\documentclass[aps,twocolumn,amsmath,amssymb,preprintnumbers,superscriptaddress,floatfix]{revtex4}
\pdfoutput=1

\usepackage{physics}

\usepackage[utf8]{inputenc}
\usepackage{newtxtext}
\usepackage[upint]{newtxmath}
\usepackage{microtype}
\usepackage{textcomp}
\usepackage{dsfont}
\usepackage{eucal}
\usepackage{siunitx}
\usepackage{soul}

\usepackage{enumerate}
\usepackage{amsfonts}
\usepackage{color}
\usepackage{soul}

\usepackage{todonotes}
\presetkeys%
    {todonotes}%
    {inline}{}

\usepackage{graphicx}

\usepackage[colorlinks,allcolors=blue]{hyperref}
\usepackage[capitalize]{cleveref} % Do we want to capitalize?
\usepackage{cleveref}

\renewcommand{\v}[1]{\boldsymbol{#1}}                           % Bold vector
\newcommand{\uv}[1]{\v{e}_{#1}}       
\definecolor{DarkBlue}{rgb}{0,0,0.80}
\definecolor{DarkRed}{rgb}{0.80,0,0}
\definecolor{Purple}{rgb}{0.55,0,0.55}
\definecolor{Purple}{rgb}{0,0,0.8}

\let\epsilon\varepsilon

\begin{document}
\title{Spin-orbit pumping}

\author{Eirik Holm Fyhn}
\affiliation{Center for Quantum Spintronics, Department of Physics, Norwegian \\ University of Science and Technology, NO-7491 Trondheim, Norway}
\author{Jacob Linder}
\affiliation{Center for Quantum Spintronics, Department of Physics, Norwegian \\ University of Science and Technology, NO-7491 Trondheim, Norway}

\date{\today}
\begin{abstract}
  \noindent We study theoretically the effect of a rotating electric field on a diffusive nanowire and find an effect that is analogous to spin pumping, which refers to the generation of spin through a rotating magnetic field.
  The electron spin couples to the electric field because the particle motion induces an effective magnetic field in its rest frame.
  In a diffusive system the velocity of the particle, and therefore also its effective magnetic field, rapidly and randomly changes direction.
  Nevertheless, we demonstrate analytically and via a physical argument why the combination of the two effects described above produces a finite magnetization along the axis of rotation.
  This manifests as a measurable spin-voltage in the range of tens of microvolts.
\end{abstract}
\maketitle
% Fakesection: Introduction
\textit{Introduction}. 
As further miniaturization of transistors becomes ever more difficult~\cite{kaul2017}, there is a pressing need for new technologies to aid or replace silicon-based information technology.
Spintronics is a candidate which has as its underlying idea to use the electron spin as an information carrier~\cite{bader2010,hirohata2020}.
This idea is promising because the spin degrees of freedom in solid state systems can potentially be manipulated in a highly energy-efficient manner.
This is important since the growing need for more computing power has significantly increased the energy consumption of information and communication technologies~\cite{jorge2020,jones2018}.
As a result, the study of spin transport and spin manipulation in low-dimensional and nanoscale devices is a growing field of research.

\begin{figure}[htpb]
  \centering
  \includegraphics[width=0.9\linewidth]{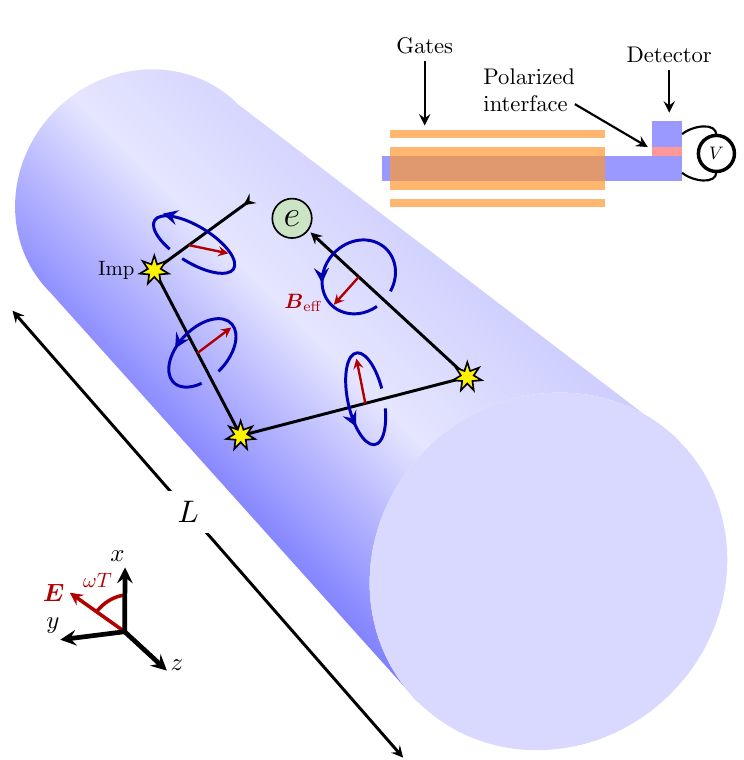}
  \caption{
    Illustration of an insulated nanowire of length $L$ subjected to an electric field, $\v E$, which rotates in the $xy$-plane with angular frequency $\omega$.
    The rotating field can be experimentally implemented using two pairs of gate voltage plates along the $x$ and $y$ axis with AC voltages and a phase-shift between the pairs, as illustrated in the top right part of the figure.
    The top right figure also illustrate a possible method of measuring the spin-accumulation through the voltage difference across a spin-polarized interface.
    Additionally, the figure illustrates the physical mechanism behind the spin pumping effect induced by the electric field.
    The effective magnetic field $\v B_\text{eff}$ in the rest frame of the electron is perpendicular to its motion in the lab frame and changes with each scattering event.
    Despite this, the projection of $B_\text{eff}$ onto the plane ($xy$) perpendicular to the nanowire length ($z$) rotates in the same direction after any scattering event, as indicated by the blue ellipses.
    This causes the spin pumping effect.
    }%
  \label{fig:sketch}
\end{figure}

One important aspect of spin manipulation is the generation of spin, which can be done by so-called spin pumping~\cite{tserkovnyak2002,tserkovnyak2005}.
This refers to the generation of spin through a precessing magnetic field.
After its discovery in ferromagnets~\cite{tserkovnyak2002}, spin pumping has been studied for a wide range of systems, such as antiferromagnets~\cite{chen2014,oyvind2017,vaidya2020}, spin-glass systems~\cite{fujimoto2020} and superconducting hybrid structures~\cite{jeon2018,kato2019,fyhn2021}.
Since the electron spin gives rise to a magnetic dipole moment, it is conceptually simplest to manipulate through magnetic fields.
However, spin also couple to electric fields since, from the perspective of a moving electron, an electric field gives rise to an effective magnetic field.
This interaction between spin and electric fields is known as spin-orbit coupling (SOC), and is the reason why electric fields play a central role in spintronics research~\cite{manchon2019,manchon2015}.
In this manuscript we investigate whether an effect analogous to spin pumping can be obtained from a time-dependent electric field through SOC.

Materials with SOC are most famously able to produce spin polarization through the spin Hall effect~\cite{sinova2015}.
This refers to how spin accumulates when a charge current is passed through, because the trajectories of electrons with opposite spins are bent in opposite directions.
This can produce a measurable spin polarization~\cite{kato2004,valenzuela2006,garlid2010}, but unlike spin pumping it requires an applied electric current.
The spin Hall effect is also widely used to detect the spin-currents produced by spin pumping~\cite{sinova2015,vaidya2020,mosendz2010}.
% This is possible because, just as an electric current generates a transverse spin-current, a spin-polarized current generates a transverse charge-current.

The prospect of spin manipulation from external electric fields is especially interesting in the context of spin-based quantum bits.
This is because magnetic fields are difficult to localize~\cite{simovic2006,torrezan2009,koppens2006} compared to their electric counterparts~\cite{nowack2007,liang2012}, something which makes individual control of spin-based quantum bits more feasible with electric fields.
Time-dependent SOC has therefore mostly been considered in quantum dots and quantum wells.
In such structures, oscillating electric fields has been studied experimentally~\cite{kato2003,nowack2007} and theoretically~\cite{rashba2003,venitucci2018,vincent2021,efros2006} with a fixed direction in space.
However, a harmonically oscillating electric field with fixed direction does not by itself break time reversal symmetry, which is necessary in order to produce magnetization.
These systems therefore require an additional static magnetic field, but an entirely electric control of the spin motion can be obtained by a rotating electric field.
This was pointed out by \citet{serebrennikov2004}, who considered an electron in a spherical potential under the influence of a rotating electric field.
More recently, \citet{entin-wohlman2020} showed that when a quantum dot subjected to a rotating electric field is placed in a junction with normal metals, the resulting time-dependent tunneling can induce a nonzero magnetization in the leads.

The prospect of spin-generation from purely electric fields from local gate electrodes is attractive also from a spintronics perspective, since such devices could be placed in close proximity to other nanoscale devices without them being affected by undesirable stray fields.
For this reason, and motivated by the success of electrical control of spin dynamics in quantum dots, we present here a study of how magnetization can be induced in diffusive nanowires by purely electrical means.
We consider an insulated wire subjected to a rotating electric field, as depicted in \cref{fig:sketch}.

In diffusive systems the physical picture is complicated by the fact that the particles rapidly change momentum direction.
This means that the effective magnetic fields also change direction frequently, as viewed from the rest frame of the particles.
Nevertheless, we find using quasiclassical Keldysh theory that a finite, time-independent magnetization is induced along the axis of rotation.
After presenting our results, we explain the physical origin of this effect.
Hence, pumping spin by rotating electric fields, which we here refer to as spin-orbit pumping, or SO pumping, can be used as an alternative to conventional spin pumping.

\begin{figure*}[htpb]
  \centering
  \includegraphics[width=0.95\linewidth]{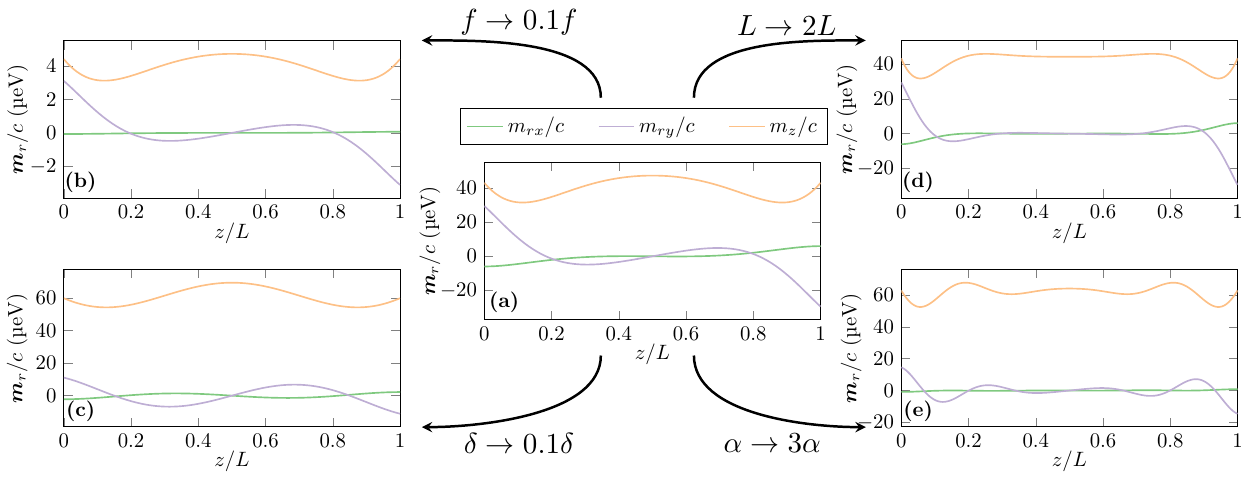}
  \caption{Spatial distribution of the components of the magnetization, $\v m_r$, for various system parameters.
    $m_{rx}$ and $m_{ry}$ are the $x$- and $y$-components of the magnetization as seen from the rotating frame of reference, while $m_z$ is the $z$-component of the magnetization and therefore the same in the rotating frame and the lab frame.
    The normalization constant is $c = \frac 1 2 g\mu_B N_0$.
    Panel \textbf{(a)} has Rashba coupling $\alpha = \SI{3e-12}{\electronvolt\meter}$, length $L = \SI{1}{\micro\meter}$, frequency $f = \SI{32}{\giga\hertz}$ and inelastic relaxation rate $\delta = \SI{0.1}{\milli\electronvolt}$.
    The remaining panels have the same parameters as $\textbf{(a)}$, except for the quantity labeled at the corresponding arrow.
    Thus, the four side panels illustrate the effect of varying $\omega$, $L$, $\delta$ and $\alpha$, respectively.
    \textbf{(b)} has $f = \SI{3.2}{\giga\hertz}$,
    \textbf{(c)} has $\delta = \SI{0.01}{\milli\electronvolt}$,
    \textbf{(d)} has $L = \SI{2}{\micro\meter}$ and
    \textbf{(e)} has $\alpha = \SI{9e-12}{\electronvolt\meter}$.
  }%
  \label{fig:mag}
\end{figure*}

\begin{figure}[htpb]
  \centering
  \includegraphics[width=1.0\linewidth]{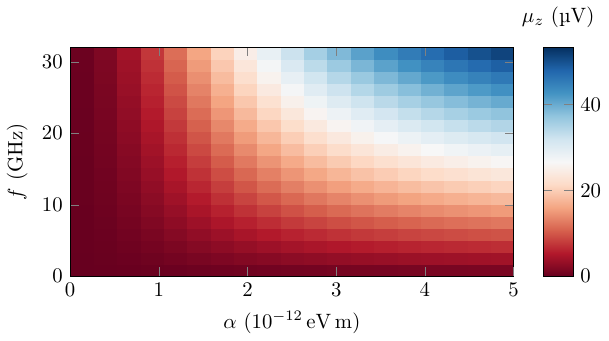}
  \caption{Spatially averaged spin-voltage as a function of Rashba coupling parameter, $\alpha$, and frequency, $f$. The spin relaxation rate is set to  $\delta = \SI{0.1}{\milli\electronvolt}$ and the length of the wire is $L = \SI{1}{\micro\meter}$.}%
  \label{fig:2d}
\end{figure}

Systems with strong atomic SOC would be advantageous in order to realize spin-orbit pumping experimentally.
The SOC in such system can, depending on the lattice symmetry, have additional static terms.
These terms will not induce spin-orbit pumping by themselves, but they can nevertheless affect the results.
We do not include such terms here, but note that it would be interesting for future work to study how the inclusion of other types of SOC can affect spin-orbit pumping.

% Fakesection: Equations and notation
\textit{Equations}.%
\label{sec:equations}
Under the assumption that the Fermi energy is the dominant energy scale and the mean free path is short, the system illustrated in \cref{fig:sketch} can be described by the quasiclassical Keldysh theory~\cite{belzig1999,rammer1986,eilenberger1968}.
Moreover, if the mean free path is much shorter than system length and the length scale associated with SOC, $1/m\alpha$, and the elastic scattering rate is much shorter than the angular frequency of the rotating electric field, the system can be classified as \emph{diffusive}.
In this case the quasiclassical Green's function $\check g_s$ solves the Usadel equation
~\cite{usadel1970,rammer1986},
\begin{equation}
  \frac{\partial\check g_s}{\partial T} + D\tilde\nabla\circ\left(\check g_s \circ \tilde\nabla\circ \check g_s\right)
  + i \comm{\check \sigma_\text{inel}}{\check g_s}_\circ = 0,
  \label{eq:usadel}
\end{equation}
Here, $T$ is time, $D$ is the diffusion constant, $\check \sigma_\text{inel}$ is the self-energy matrix from inelastic relaxation processes~\cite{virtanen2010} and $\tilde\nabla$ is the covariant derivative which includes the spin-orbit coupling.
Moreover, the circle-product is
\begin{equation}
  a\circ b = \exp\left(\frac i 2 \partial_\varepsilon^a\partial_T^b - \frac i 2 \partial_T^a\partial_\varepsilon^b\right)ab,
  \label{eq:circ}
\end{equation}
where $\varepsilon$ is energy.
The time-varying electric field will generally also induce a magnetic field, but we find that this is negligible compared to the effective magnetic field felt by the moving particles due to the electric field.

The circle-product makes the Usadel equation difficult to solve in time-dependent situations, but in this case it can be simplified by a Fourier transform in energy~\cite{fyhn2021_time}.
From this we can find an equation for the magnetization,
\begin{equation}
  \v m = \frac{g \mu_B N_0}{16} \int_{-\infty}^{\infty}\dd{\varepsilon} \Tr(\v\sigma \check g_s^K),
\end{equation}
where the superscript $K$ denotes the Keldysh part, $\v\sigma$ is the vector of Pauli matrices, $g$ is the Landé $g$-factor, $N_0$ is the density of states at the Fermi energy, $\mu_B$ is the Bohr magneton.
We find that $\v m$ solves
\begin{multline}
  \frac{\partial\v m}{\partial T} - D\frac{\partial^2\v m}{\partial z^2} + 2\delta \v m=  
  4D\v A \times \pdv{\v m}{z} + 4D \v A \times (\v A \times \v m)
  \\- g \mu_B N_0 D \v A \times \frac{\partial\v A}{\partial T},
  \label{eq:magnetization}
\end{multline}
as shown in the supplementary material~\cite{supplementary,houzet2008}.
Here, $\v A$ comes from the spin-orbit coupling and is given by
\begin{equation}
  \v A = m\alpha \uv E \times \uv z,
  \label{eq:soc}
\end{equation}
where $\alpha$ is the Rashba coupling, $m$ is the effective mass, $\uv z$ is the unit vector in the $z$-direction and $\uv E$ is the unit vector pointing in the direction of the electric field, which has been assumed to be uniform in space.
Finally, $\delta$ is an effective parameter describing the spin relaxation rate from sources other than spin-orbit coupling, such as inelastic phonon scattering.
The spin relaxation rate described by $\delta$ is assumed independent of spin-direction.
Moreover, \cref{eq:soc} already contains a relaxation term that depends on the spin-direction due to $\v{A}$ which we comment on below.

The left hand side of \cref{eq:magnetization} describes diffusion and the spin relaxation in the absence of SOC, while the right hand side is the effect of the SOC.
The first term on the right side describes spin precession of the diffusion current and the second term is spin relaxation due to the Dyakonov-Perel mechanism~\cite{dyakonov1972}.
This relaxation comes from the randomization of spin precession angles caused by elastic scattering at non-magnetic impurities.
The third and final term is the source term coming from the time-dependence of $\v A$.
It is this term which makes SOC capable of producing spin in diffusive systems.
We can from this term immediately see that a time-varying electric field with fixed direction will not generate spin, since $\v A\times \partial \v A/\partial T = 0$ in that case.

\Cref{eq:magnetization} must be accompanied by boundary conditions. 
For simplicity we choose insulating boundaries, which means that the particle flux across the interfaces at $z=0$ and $z=L$ must be zero.
In the diffusive limit of the quasiclassical Green's function formalism, the relevant boundary condition is known as the Kupriyanov-Lukichev boundary condition~\cite{KL1988}.
From this we find that
\begin{equation}
  \frac{\partial \v m}{\partial z} + 2 \v A \times \v m - \frac 1 2 g \mu_B  N_0 \frac{\partial \v A}{\partial T} = 0
\end{equation}
at $z=0$ and $z=L$, as shown in the supplementary material~\cite{supplementary,chandrasekhar2008,ali2018}.

\Cref{eq:magnetization} can be solved for times long after the rotating electric field has been turned on by looking for a stationary solution in the rotating reference frame.
This is because all solutions converge to this unique stationary solution, as we prove in the supplementary material~\cite{supplementary}.
In the rotating reference frame, the electric field is time-independent and the magnetization along the $z$-direction is the same as in the lab frame.
Converting the equations into the rotating frame can be done by inserting the rotation matrix $R(\omega T)$ which satisfies
\begin{equation}
  \v A = \mqty(\cos(\omega T) & \sin(\omega T) & 0 \\ -\sin(\omega T) & \cos(\omega T) & 0 \\ 0 & 0 & 1) \v A_0 = R(\omega T) \v A_0,
  \label{eq:rot}
\end{equation}
where $\v A_0$ is constant in time.
We choose $\v A_0 = \abs{\v A_0} \uv x$.
To write \cref{eq:magnetization} in the rotating system, we write $\v m = R(\omega T)\v m_r$ and use the relations $[R(\theta)\v u]\times[R(\theta)\v v] = R(\theta) \v u \times \v v$ and
\begin{equation}
  \frac{\partial}{\partial T} R(\omega T) \v u = R(\omega T)\frac{\partial \v u}{\partial T}
  + R(\omega T) \left[\v u \times \omega\uv z  \right].
\end{equation}
The equation for the magnetization in the rotating frame is therefore
\begin{multline}
  D\frac{\partial^2\v m_r}{\partial z^2} -2\delta \v m_r +
  \v \Omega \times \v m_r
  +4D\v A_0 \times \frac{\partial\v m_r}{\partial z} 
  \\+ 4D \v A_0 \times \left[\v A_0 \times \left(\v m_r - \frac{g\mu_B N_0}{4}\v \Omega\right)\right]=0,
  \label{eq:magnetizationRot}
\end{multline}
where we have used that $\v m_r$ is independent of time and $\v \Omega = \omega \uv z$.
The boundary condition in the rotating frame is
\begin{align}
    \frac{\partial \v m_r}{\partial z} + 2 \v A_0 \times \left(\v m_r - \frac{g\mu_B N_0}{4} \v \Omega\right) &= 0,
    \label{eq:bcRot}
\end{align}
at $z = 0$ and $z = L$.

The magnetization is measurable through the so-called spin-voltage,
\begin{equation}
  \mu_z = \frac{m_z}{\frac 1 2 g\mu_B N_0 \abs{e}}.
\end{equation}
If we connect the nanowire to a detector electrode through an interface with polarization $P$ along the $z$-direction, as illustrated in \cref{fig:sketch}, then $P\mu_z$ is the voltage difference between the nanowire and ferromagnet in the absence of electric current~\cite{silsbee1980,johnson1985,tombros2007,poli2008,silaev2015,heikkila2019}.
This is shown in the supplementary material~\cite{supplementary,bergeret2012,ali2017,hugdal2017}.

% Fakesection: Results
\textit{Results}.
We solve \cref{eq:magnetizationRot,eq:bcRot} numerically by using the finite element method.
\Cref{fig:mag} shows the resulting spatial distribution of $\v m_r$ for various system parameters and \cref{fig:2d} shows the spatially averaged spin-voltage as a function of the Rashba coupling parameter and frequency, $f = \omega/2\pi$.
We have used $m = 0.1m_e$, where $m_e$ is the electron mass and $D = \SI{e2}{\centi\meter\squared\per\second}$.
Note that the $z$-component of the magnetization is equal in the rotating frame and lab-frame, so $m_z$ in \cref{fig:mag} is static and equal in both frames.

\Cref{fig:mag} shows that the magnetization has non-zero components in the $x$- and $y$-direction that are antisymmetric around the middle of the wire.
In the rotating frame, magnetization along the $y$-direction is induced at the boundaries, as can be seen from \cref{eq:bcRot}.
This magnetization is rotated around the $z$-axis from the term $\v\Omega \times \v m_r$ in \cref{eq:magnetizationRot}, and it is rotated into the $z$-component because of $4D\v A_0 \times \partial \v m_r / \partial z$.
The former comes from the effective magnetic field present in the rotating frame and the latter is the spin precession of the diffusion current from SOC.
We can see the spin rotation effect of $\v\Omega$ by comparing panels (a) and (b) in \cref{fig:mag}.
From this we see that when $f$ is decreased from $\SI{32}{\giga\hertz}$ to $\SI{3.2}{\giga\hertz}$, $m_z$ and $m_{ry}$ are also scaled by a factor $1/10$.
This is reasonable since the source term in the magnetization equation is proportional to $f$.
The $x$-component, $m_{rx}$, on the other hand, is reduced much more in \cref{fig:mag}~(b), which is expected since the rotation from $m_{ry}$, coming from the term $\v\Omega \times \v m_r$, is much less.

Comparing \cref{fig:mag}~(a) to \cref{fig:mag}~(c) and (e) shows the effect of decreasing the inelastic relaxation and increasing the Rashba coupling, respectively.
In both cases the ratio between the spin generation from SOC and the inelastic spin relaxation is increased.
As a result, the magnetization along the $z$-axis is increased.
From \cref{eq:bcRot} we see $\abs{\partial m_{ry}/\partial z}$ gets smaller at the boundaries when $m_z$ is larger.
This is reflected in the smaller $y$-component in \cref{fig:mag}~(c) and (e).
Finally, unlike \cref{fig:mag}~(c), \cref{fig:mag}~(e) has more rapid oscillations in $m_z$ and $m_{ry}$.
This is expected since increasing $\alpha$ not only increases the spin generation, but also the spin precession associated with SOC.

From \cref{fig:mag,fig:2d} we see that the rotating electric field can produce a spin-voltage of tens of microvolt with the parameters used here.
This is our main result and shows that spin-orbit pumping is be capable of producing a measurable magnetization.
We propose that the SO pumping effect can be understood in terms of normal spin pumping from the effective magnetic field in the reference frames of the moving charge carriers.
Consider a particle with velocity $\v v = (v_x, v_y, v_z)$ moving in the effective electric field $\v E = E(\cos(\omega T), \sin(\omega T), 0)$.
The effective magnetic field is obtained via a Lorentz-transformation:
\begin{equation}
  \v B_\text{eff} = \mqty(v_z\sin(\omega T) \\ -v_z\cos(\omega T) \\ v_y \cos(\omega T) - v_x\sin(\omega T))E.
\end{equation}
This effective field rotates in an elliptical way around an axis.
Although the direction of this axis changes with the particle velocity, its component along the $z$-axis is always of the same sign.
This is illustrated in \cref{fig:sketch} and can be most easily seen by noting that the projection of $\v B_\text{eff}$ onto the $xy$-plane always rotates counter-clockwise when $\omega > 0$ and clockwise when $\omega < 0$.
Since it is known from normal spin pumping that a rotating magnetic field induces a magnetization along the axis of rotation, this explains why a rotating electric field can generate a magnetization in the $z$-direction.
Summarized, the physical picture of SO pumping in diffusive systems is as follows.
With each elastic scattering, the spin precession axis jumps to a new direction.
This randomizes the spin over time and gives rise to a spin relaxation.
This is just the normal Dyakonov-Perel mechanism.
However, since the electric field rotates, the spin precession axis also rotates between scatterings.
Since this rotation is always in the same direction around the $z$-axis it gives rise to a net spin-accumulation polarized in the $z$-direction.
The equivalence in the quasiclassical theory between SOC and the effective magnetic field $\v B_\text{eff}$ is shown explicitly in the supplementary material~\cite{supplementary}.

One difference from normal spin pumping is that a rotating electric field both generate and dissipate spin because of the Dyakonov-Perel mechanism.
Thus, by increasing the electric field strength, both spin generation and spin relaxation is increased.
When Dyakonov-Perel relaxation is the dominant spin relaxation mechanism, we can see from \cref{eq:magnetizationRot,eq:bcRot} that the spin generation and spin relaxation mechanisms equalize when $\v m_r = g \mu_B N_0 \v\Omega /4$.
This can be seen from the fact that $\v m_r = g \mu_B N_0 \v\Omega /4$ solves \cref{eq:magnetizationRot,eq:bcRot} when $\delta = 0$.
Thus, SO pumping in diffusive systems can at most produce as spin-voltage of $\mu_z = \omega/2\abs{e} \approx 2\times(f/\si{\giga\hertz})~\si{\micro\volt}$.
However, in the presence of other spin relaxation mechanisms, the observed spin-voltage will be less, as is the case in \cref{fig:mag,fig:2d}. 

Based on the physical picture of SO pumping as the cumulative effect of normal spin pumping from the rotating effective magnetic field observed between each scattering, it is clear that scattering processes work to reduce the SO pumping effect.
It would therefore be of interest to study rotating electric fields in clean, ballistic systems to see if the SO pumping effect can be enhanced in such systems.
We leave this for future work.

% Fakesection: Conclusion
\textit{Conclusion}.
We have found using quasiclassical Keldysh theory that a rotating electric field can induce a magnetization and a measurable spin-voltage of tens of $\si{\micro\volt}$.
This spin-orbit pumping can be understood as a spin pumping from the effective magnetic field in the rest frame of the moving particles.
This is because, despite the jumps occurring at each scattering event, the projection of the effective magnetic field onto the plane in which the electric field is applied always rotates in the same direction.
Obtaining a spin-voltage above $\SI{10}{\micro\volt}$ with the material parameters used here requires a Rashba coupling of $\SI{e-12}{\electronvolt\meter}$.
Rashba coupling strengths of this magnitude has been obtained experimentally at temperatures below $\SI{15}{\kelvin}$ in nanowires with applied electric fields~\cite{liang2012,takase2017}.
One reason for this requirement is that spin relaxation, both from inelastic relaxation and from SOC through Dyakonov-Perel relaxation, inhibits spin-orbit pumping.
Thus, it would be of interest to study rotating electric fields in clean, ballistic nanowires to see whether the spin-orbit pumping effect is stronger in such systems.
Nevertheless, the findings presented here shows that spin-orbit pumping should be capable of producing an experimentally observable magnetization even in diffusive systems.

% Fakesection: Acknowledgements
\begin{acknowledgments}
This work was supported by the Research Council of Norway through grant 240806, and its Centres of Excellence funding scheme grant 262633 ``\emph{QuSpin}''. J. L. also acknowledge funding from the NV-faculty at the Norwegian University of Science and Technology. 
\end{acknowledgments}

% Fakesection: Bibliogaphy
% \clearpage
\bibliography{bibliography}

\end{document}

\title{Supplementary: Spin-orbit-pumping}

\author{Eirik Holm Fyhn}
\affiliation{Center for Quantum Spintronics, Department of Physics, Norwegian \\ University of Science and Technology, NO-7491 Trondheim, Norway}
\author{Jacob Linder}
\affiliation{Center for Quantum Spintronics, Department of Physics, Norwegian \\ University of Science and Technology, NO-7491 Trondheim, Norway}

\date{\today}
\maketitle
\section{Derivation of magnetization equation}%
\label{sec:kinetic_equations}
When a charged particle with mass $m$ moves with momentum $\v p$ in an electric field $\v E = E \uv E$, it will feel an effective magnetic field, $\v B_\text{eff} = \v E \times \v p/m$ (in natural units), and therefore also an effective Zeeman energy
\begin{equation}
  \mathcal H_\text{SOC} = \alpha \left(\v \sigma \times \v p\right) \cdot \uv E,
  \label{eq:soc_1}
\end{equation}
where $\v \sigma$ is the vector of Pauli-matrices, such that $\v\sigma /2$ is the spin operator, and $\alpha = g\mu_B E/2m$ is a parameter giving the strength of the spin-orbit coupling (SOC).
Here, $g$ is the Landé $g$-factor and $\mu_B$ is the Bohr magneton.
In realistic system the SOC can be more complicated and depend on the crystal structure and atomic potential. 
Nevertheless, the spin-orbit effect can often be approximated by an Hamiltonian on the form of \cref{eq:soc_1} where $\uv E$ is the unit vector in the direction of the external electric field and $\alpha$ is an effective parameter called the Rashba coupling~\cite{manchon2015}.
Here we use this Rashba form for the spin-orbit coupling and keep $\alpha$ as a free parameter.

The system under consideration can be treated quasiclassically if the material has a well-defined Fermi surface and the Fermi wavelength is much shorter than all other relevant length-scales, such as the mean free path, system length and the length scale associated with SOC, $1/m\alpha$.
In this case, the system can be described by quasiclassical Green's functions, which can be collected in a $4\times 4$ matrix as
\begin{equation}
  \check g = \mqty(g^R & g^K \\ 0 & g^A),
\end{equation}
where $g^R$, $g^A$ and $g^K$ are the retarded, advanced and Keldysh quasiclassical Green's functions, respectively.
These are normalized such that $\check g\circ \check g = 1$ and solve the Eilenberger equation~\cite{eilenberger1968,belzig1999},
\begin{equation}
  \frac{\partial\check g}{\partial T} + \v v_F \cdot \tilde\nabla\circ \check g -i \comm{\check\sigma_\text{inel} - \frac{i}{2\tau}\check g_s}{\check g}_\circ = 0,
  \label{eq:eilenberger}
\end{equation}
where $\v v_F$ is the Fermi velocity, $T$ is time, $\tau$ is the elastic impurity scattering time, $\check g_s$ is the isotropic part of the quasiclassical Green's function and $\check\sigma_\text{inel}$ is the self-energy matrix from inelastic relaxation processes.
Moreover, the circle-product is
\begin{equation}
  a\circ b = \exp\left(\frac i 2 \partial_\varepsilon^a\partial_T^b - \frac i 2 \partial_T^a\partial_\varepsilon^b\right)ab,
  \label{eq:circ}
\end{equation}
where $\varepsilon$ is energy and the covariant derivative is
\begin{equation}
  \tilde\nabla\circ \check g = \nabla \check g - i \left(\v a\circ \check g - \check g\circ \v a\right),
\end{equation}
where $\v a$ depend on the Rashba coupling through
\begin{equation}
  \v a = m\alpha \v \sigma \times \uv E.
\end{equation}
We include inelastic relaxation through the relaxation time approximation~\cite{virtanen2010}.
In this approximation the relaxation rate is given by $\delta$ and we assume that it is isotropic in spin-space.
In this case, $\sigma_\text{inel}^R = -\sigma_\text{inel}^A = i\delta$ and $\sigma_\text{inel}^K = 2i\delta h_\text{eq}$, where $h_\text{eq}$ is the equilibrium distribution function which the system relaxes towards.
At inverse temperature $\beta$ and electrochemical potential $V$, this is $h_\text{eq}(\varepsilon) = \tanh[\beta(\varepsilon - eV)/2]$, where $e$ is the electron charge.
Generally, SOC also gives rise to a term in the self-energy, which is $\sigma^R_\text{SOC} = \sigma^A_\text{SOC} = -m\alpha^2$.
However, since we assume that $\alpha$ is constant in time, $\check\sigma_\text{SOC}\circ \check g = \check g\circ\check\sigma_\text{SOC}$, and $\check\sigma_\text{SOC}$ disappears from \cref{eq:eilenberger}.

We can see how the effective magnetic field discussed in the main text enters the quasiclassical framework by noting that if $\v B_\text{eff} = (2/g\mu_B)\alpha m \uv E \times \v v_F$, we can rewrite \cref{eq:eilenberger} as
\begin{equation}
  \frac{\partial\check g}{\partial T} + \v v_F \cdot \nabla \check g -i \comm{\frac 1 2 g \mu_B \v B_\text{eff}\cdot\v\sigma - \frac{i}{2\tau}\check g_s + \check\sigma_\text{inel}}{\check g}_\circ = 0,
  \label{eq:eilenberger2}
\end{equation}
where $\v \sigma$ is the vector of Pauli matrices.
The term $\frac 1 2 g \mu_B \v B_\text{eff}\cdot\v\sigma$ is exactly how the Zeeman energy from a magnetic field $\v B_\text{eff}$ would enter as a self-energy the Eilenberger equation.
In other words, the spin-orbit coupling present in the covariant derivative is functionally equivalent to an external magnetic field $\v B_\text{eff}$.
This is in accordance with the physical picture of spin-orbit pumping as an effective spin-pumping from $\v B_\text{eff}$, as discussed in the main text.

The isotropic part of $\check g$ dominates when the impurity scattering time $\tau$ is small, such that the elastic impurity scattering term $-i\check g_s/2\tau$ is large.
In particular, if the scattering time $\tau$ is much smaller than the inelastic scattering time, $1/\delta$, and the rate at which the Green's function changes, which in this case is given by the frequency of the rotating electric field, $\omega$, and if the corresponding mean free path $l_\text{mfp} = v_F\tau$ is much smaller than the system length as well as the length scale associated with SOC, $1/m\alpha$, then the Eilenberger equation reduces to the Usadel equation~\cite{usadel1970,rammer1986},
\begin{equation}
  \frac{\partial\check g_s}{\partial T} + D\tilde\nabla\circ\left(\check g_s \circ \tilde\nabla\circ \check g_s\right)
  + i \comm{\check \sigma_\text{inel}}{\check g_s}_\circ = 0,
  \label{eq:usadel}
\end{equation}
where $D$ is the diffusion constant.
In the non-superconducting case considered here, the equations are considerably simplified by the fact that the retarded and advanced are simply proportional to the unitary matrix.
In this case we have $g_s^R = I_2$, $g_s^A = -I_2$, and $g_s^K = 2h$, where $I_2$ is the $2\times 2$ identity matrix and $h$ is the distribution function.
From \cref{eq:usadel,eq:circ} and the relaxation time approximation we get that the distribution function solves
\begin{equation}
  \frac{\partial h}{\partial T} - D\tilde\nabla\circ\tilde\nabla\circ h + 2\delta\left(h - h_\text{eq}\right) = 0.
  \label{eq:distr_energy}
\end{equation}

The circle-products can be removed either by a unitary transformation~\cite{houzet2008} or a Fourier transform in energy~\cite{fyhn2021_time}.
Here we choose the latter and use capital letters to denote Fourier transforms,
\begin{equation}
  H(t, T, z) = \mathcal F\{h\}(t, T, z) = \frac{1}{2\pi} \int_{-\infty}^\infty\dd{\varepsilon} h(\varepsilon, T, z) \me{-i\varepsilon t},
\end{equation}
where $z\in (0,L)$ is the position along the wire.
Next, we define the $z$-component of $\v a$ to be $a_z = \v A \cdot \v \sigma$
and $H = H_0 + \v H \cdot \v \sigma$.
The Fourier transform is useful because
\begin{equation}
  \mathcal F\left\{\comm{\v A \cdot \v \sigma}{h}_\circ\right\}(t, T, z)
  = \v A(T+t/2)\cdot\v \sigma H(t,T,z) - H(t,T,z)\v A(T-t/2)\cdot\v \sigma.
\end{equation}
We can use that for the Pauli matrices we have $\sigma_i\sigma_j = \delta_{ij} + i\varepsilon_{ijk}\sigma_k$, where $\varepsilon_{ijk}$ is the Levi-Civita symbol, and define $\v A_\pm(T,t) = \v A(T+t/2)\pm \v A(T-t/2)$ to get
\begin{equation}
  \mathcal F\left\{\comm{\v A \cdot \v \sigma}{h}_\circ\right\}
  = \v A_- \cdot \v H + \left(\v A_- H_0 + i \v A_+ \times \v H\right) \cdot \v \sigma.
\end{equation}
Using this we get that
\begin{equation}
  \frac 1 2 \Tr[\v\sigma \mathcal F\left\{ \tilde\nabla\circ\tilde\nabla\circ h\right\}]
  = \frac{\partial^2 \v H}{\partial z^2}
  - 2i \v A_- \frac{\partial H_0}{\partial z} 
  + 2 \v A_+ \times \frac{\partial\v H}{\partial z}
  - \v A_-\left(\v A_-\cdot \v H\right)
  - i \v A_+ \times \v A_- H_0 + \v A_+ \times \left(\v A_+ \times \v H\right).
\end{equation}
Thus, by Fourier transforming \cref{eq:distr_energy}, multiplying it with $\v \sigma/2$ and taking the trace, we get that
\begin{equation}
  \frac{\partial \v H}{\partial T} - D \frac{\partial^2 \v H}{\partial z^2} 
  + 2iD \v A_-\frac{\partial H_0}{\partial z}
  - 2D\v A_+ \times \frac{\partial\v H}{\partial z}
  + D\v A_-\left(\v A_-\cdot \v H\right)
  + iD\v A_+ \times \v A_- H_0 - D\v A_+ \times \left(\v A_+ \times \v H\right)
  + 2\delta \v H = 0.
  \label{eq:distr_t}
\end{equation}

In the quasiclassical framework, magnetization is given by $\v m = \frac 1 2 g \mu_B N_0 \pi\lim_{t\to 0}\Re(\v H)$ in the absence of an exchange field.
All we need to get an equation for $\v m$ is therefore to take the limit $t\to 0$ of \cref{eq:distr_t}.
This requires some care, since $H_0$ is asymptotic to $H_\text{eq}$ as $t\to 0$, and $H_\text{eq} = -i/[\beta\sinh(\pi t/\beta)]$ diverges as $t \to 0$. 
Nevertheless, the limit in \cref{eq:distr_t} is well-defined since $H_0$ only occurs multiplied by $\v A_-$, which goes to $0$ as $t\to 0$.

First, we show that $H_0$ is asymptotic to $H_\text{eq}$.
This is a consequence of the fact that the physics happens close to the Fermi-surface, meaning that for energies far away from the Fermi-surface the states are either fully occupied ($h_0(\varepsilon) = -1$) or entirely empty ($h_0(\varepsilon) = 1$).
That is, $\abs{\varepsilon} \gg 1 \implies h_0(\varepsilon)  \approx h_\text{eq}(\varepsilon)\approx \varepsilon/\abs{\varepsilon}$.
Thus,
\begin{equation}
  \lim_{t\to 0} i\pi t H_0 = \lim_{t\to 0} \frac{1}{2}it\mathcal F\{h_0\} = \lim_{t\to 0}\frac 1 2 \mathcal F\left\{ \frac{\partial h_0}{\partial \varepsilon} \right\} 
  = \frac 1 2 \int_{-\infty}^{\infty} \dd{\varepsilon} \frac{\partial h_0}{\partial \varepsilon} = 1.
\end{equation}
Hence, $H_0 \sim 1/i\pi t$ as $t\to 0$.
Finally, using that $\lim_{t\to 0}\v A_+(T, t) = 2 \v A(T)$, $\lim_{t\to 0}\v A_-(T,t)/t = \partial \v A(T)/\partial T$, we get the equation for the equation for the magnetization by taking the limit $t\to 0$ of \cref{eq:distr_t} and multiplying it with $\frac 1 2 g \mu_B N_0 \pi$.
This yields
\begin{align}
  \frac{\partial\v m}{\partial T} - D\frac{\partial^2\v m}{\partial z^2} + 2\delta \v m=  
  4D\v A \times \frac{\partial \v m}{\partial z} + 4D \v A \times (\v A \times \v m)
  - g \mu_B N_0 D \v A \times \frac{\partial\v A}{\partial T},
  \label{eq:magnetization}
\end{align}
which is the same as Eq. (1) in the main text.

\section{Boundary condition}%
\label{sec:boundary_condition}
Quasiclassical theory is not valid across interfaces because the relevant length scale is short.
Consequently, the quasiclassical Green's function is generally not continuous.
Instead, the quasiclassical Green's functions in neighbouring materials are connected through boundary conditions.
The boundary conditions express the so-called matrix current,
\begin{equation}
  \check{\v I}(\v R, \varepsilon, T) = \int \dd{\Omega} \v v_F \check g(\v v_F, \v R, \varepsilon, T),
\end{equation}
where the integral goes over all directions of the Fermi velocity, in terms of the propagators on both sides of the interface.
Only the Keldysh-component is nonzero in the our case.
The matrix current contain in its Keldysh-component both the electrical current and the spin-current, as well as the heat-current and so-called spin-heat-current~\cite{chandrasekhar2008,ali2018}.
There should be no current across insulating, spin-inactive interfaces, and so in this case
\begin{equation}
  \uv n \cdot \check{\v I}^K = -2D\uv n \cdot \tilde\nabla\circ h = 0,
\end{equation}
where $\uv n$ is the unit normal vector pointing out of the interface.
Here, we have used that in the diffusive limit $\check{\v I} = -D\left(\check g_s \circ \tilde\nabla\circ \check g_s\right)$.

More generally, one can use the Kupriyanov-Lukichev boundary condition~\cite{KL1988},
\begin{equation}
  N_{0i}D_i\uv n \cdot \left(\check g_{si} \circ \tilde\nabla\circ \check g_{si}\right)^K = \frac{\zeta}{2} \comm{\check g_{si}}{\check g_{sj}}_\circ^K,
  \label{eq:KL}
\end{equation}
which is valid for low-transparency tunneling interfaces with no spin-active properties.
Here, $N_{0i}$ and $D_i$ is the density of states at the Fermi surface and diffusion constant in material $i$, respectively, $\uv n$ is the unit vector orthogonal to the interface and pointing from material $i$ to material $j$, and $\zeta$ is the conductance across the interface.
We can rewrite \cref{eq:KL} to
\begin{equation}
  \uv n\cdot \tilde\nabla\circ h_i = \frac{\zeta}{N_{0i}D_i} \left(h_j - h_i\right).
\end{equation}
We assume that the nanowire is insulated, so we set $\zeta = 0$.
Taking the Fourier transform, multiplying with $\v\sigma/2$, taking the trace and assuming that $\uv n$ is in the $z$-direction, we get that
\begin{equation}
  \frac{\partial\v H}{\partial z} + \v A_+ \times \v H - i H_0 \v A_- = 0,
\end{equation}
where we have dropped the subscript $i$.
Again multiplying by $\frac 1 2 g \mu_B N_0 \pi$ and taking the limit $t \to 0$ we get
\begin{equation}
  \frac{\partial\v m}{\partial z} + 2\v A \times \v m -  \frac 1 2 g \mu_B N_0\frac{\partial \v A}{\partial T} = 0,
  \label{eq:boundarycond}
\end{equation}
which is the boundary condition used in the main text.

\section{Convergence to the stationary solution and its uniqueness}%
\label{sec:convergence_to_the_stationary_solution_and_its_uniqueness}
In this section we show that regardless of initial condition, all solutions of \cref{eq:magnetization} together with the boundary condition, \cref{eq:boundarycond}, converge to the solution we present in the main manuscript for times long after the electric field has been turned on.
In so doing, we also show that this solution is unique.

We start from the equation in the rotating fram, as derived in the main manuscript,
\begin{align}
  -\frac{\partial \v m_r}{\partial T} + D\frac{\partial^2 \v m_r}{\partial z^2} - 2\delta \v m_r + \v \Omega \times \v m_r + 4D\v A_0 \times \frac{\partial \v m_r}{\partial \v z} 
  + 4 D \v A_0\times \left[\v A_0 \times \left(\v m_r - \frac{g \mu_B N_0}{4}\v \Omega\right)\right] = 0,
  \label{eq:1}
\end{align}
and the boundary condition,
\begin{equation}
  \frac{\partial \v m_r}{\partial z} + 2 \v A_0 \times \left(\v m_r - \frac{g \mu_B N_0}{4}\v \Omega\right) = \v 0.
  \label{eq:2}
\end{equation}
Let the stationary solution found in the main text be $\v u$.
What we want to show is that all solutions converge to $\v u$ as time goes to infinity.
Let $\v v$ be any other solution to \cref{eq:1,eq:2}.
Next, define $\v w = \v u - \v v$.
We want to show that $\v w$ must go to zero.
Inserting $\v w$ in the equation above, we get
\begin{align}
  -\frac{\partial \v w}{\partial T} &+ D\frac{\partial^2 \v w}{\partial z^2} - 2\delta \v w + \v \Omega \times \v w + 4D\v A_0 \times \frac{\partial \v w}{\partial \v z}
  + 4 D \v A_0\times \left(\v A_0 \times \v w\right) = 0,
  \\
  \frac{\partial \v w}{\partial z} &+ 2 \v A_0 \times \v w  = 0 \qquad \text{at $z = 0$ and $z = L$.}
\end{align}
Next, we introduce the rotation matrix
\begin{equation}
  R = \begin{pmatrix}
    1 & & \\
      & \cos(2A_0 z) & -\sin(2A_0 z) \\
      & \sin(2A_0 z) & \cos(2A_0 z)
  \end{pmatrix},
\end{equation}
where $A_0 = \lvert\v A_0 \rvert$, and define $\tilde{\v w} = R\v w$.
Recall that $\v A_0$ points in the $x$-direction, so
\begin{align}
  -\frac{\partial \tilde{\v w}}{\partial T} &+ D\frac{\partial^2 \tilde{\v w}}{\partial z^2} - 2\delta \tilde{\v w} + (R\v \Omega) \times \tilde{\v w}  = 0,
  \\
  \frac{\partial\tilde{\v w}}{\partial z} &= 0\qquad \text{at $z = 0$ and $z = L$,}
\end{align}
where we used $R[\partial_z \v w + 2\v A_0 \times \v w] = \partial_z(R\v w)$ and $R(\v a\times \v b) = (R\v a)\times (R\v b)$, for any vectors $\v a$ and $\v b$.
Next, take the dot product of these equations with $\tilde{\v w}$ and use $(\v a \times \v b)\cdot \v b = 0$ to obtain
\begin{align}
  -\frac 1 2 \frac{\partial \tilde w^2}{\partial T} &+ \frac D 2 \frac{\partial^2 \tilde w^2}{\partial z^2} = D\left\lvert \frac{\partial \tilde{\v w}}{\partial z} \right\rvert^2 + 2\delta \tilde w^2,
  \label{eq:omegasquared}
  \\
  \frac{\partial\tilde w^2}{\partial z} &= 0.
\end{align}
Finally, we define $W^2 = \int_0^L dz \tilde w^2$ and integrate \cref{eq:omegasquared} to obtain
\begin{equation}
  \frac{\partial W^2}{\partial T} = -2\int_0^L dz \left(D\left\lvert \frac{\partial \tilde{\v w}}{\partial z} \right\rvert^2 + 2\delta \tilde w^2\right).
\end{equation}
The right hand side is negative as long as $W^2 \neq 0$.
Moreover, since $W^2 \geq 0$, we see that $W^2 \to 0$ as $T \to \infty$.
Hence, the stationary solution is unique and all solutions converge to the stationary solution independently of initial condition.

\section{Detector setup}%
\label{sec:detector_setup}
In order to detect the spin-voltage one can connect the nanowire to a detector electrode through a polarized tunneling boundary~\cite{silsbee1980,johnson1985,tombros2007,poli2008,silaev2015,heikkila2019}.
The detector can be a normal metal or ferromagnet.
In the case of a normal metal, a spin-polarized interface can for instance be achieved by inserting a thin ferromagnetic insulator between the nanowire and detector electrode.
In the following we show how the voltage difference between the nanowire and electrode can be used to determine the spin-voltage, as discussed in the main text.

\begin{figure}[]
  \centering
  \includegraphics[width=0.3\linewidth]{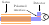}
  \caption{Illustration of the system under consideration together with the proposed detection setup. The gates encapsulating the nanowire produce a magnetization inside the nanowire through spin-orbit pumping. The resulting spin-voltage can in turn be detected through the voltage difference $V$ between the detector and the nanowire, given that the interface is polarized.}%
  \label{fig:sketch_setup}
\end{figure}

Consider the detector setup illustrated in \cref{fig:sketch_setup}.
The detector is connected to the nanowire through a polarized interface and forms an open circuit.
Charge, unlike spin, is conserved inside the detector.
Hence, the charge current into the electrode must be zero in the stationary state.
Assuming that the interface is polarized in the $z$-direction, the charge current into the detector can generally be written
\begin{equation}
  I_\text{det} = G_\up (V^\text{det} - \mu_z) + G_\dn (V^\text{det} + \mu_z),
  \label{eq:curr_det_1}
\end{equation}
where $G_\up$ and $G_\dn$ is the conductances for electrons with spin up and spin down, respectively.
The spin-voltage inside the nanowire is $\mu_z$ and the voltage difference between the detector and nanowire is $V^\text{det}$.
It is assumed that the spin-diffusion length inside the detector is short, such that the spin-voltage inside the detector is much smaller than $V^\text{det}$.
The voltage inside the detector electrode will stabilize at the value satisfying $I_\text{det} = 0$, which we find from \cref{eq:curr_det_1} happens at
\begin{equation}
  V^\text{det} = \frac{G_\up - G_\dn}{G_\up + G_\dn} \mu_z = P \mu_z,
\end{equation}
where we have inserted the polarization $P = (G_\up - G_\dn)/(G_\up + G_\dn)$.

We can derive \cref{eq:curr_det_1} more rigorously in the quasiclassical theory.
In this way we can take into account the finite spin-voltage which will also be induced in the detector.
In order to capture spin-active tunneling boundaries in the quasiclassical framework, we can use a generalization of the Kupriyanov-Lukichev condition~\cite{bergeret2012,ali2017},
\begin{equation}
  N_{0i}D_i\uv n \cdot \left(\check g_{si} \circ \tilde\nabla\circ \check g_{si}\right)^K = \frac{\zeta}{2} \comm{\check g_{si}}{(t + u\sigma_z)\check g_{sj}(t + u\sigma_z)}_\circ^K - i\frac{G_\phi}{2}\comm{g_{si}^K}{\sigma_z}_\circ,
  \label{eq:KL_polarized}
\end{equation}
where we have assumed that the interface is polarized in the $z$-direction, $t = \sqrt{(1+\sqrt{1-P^2})/2}$, $u = \sqrt{(1-\sqrt{1-P^2})/2}$ and $G_\phi$ is the spin-mixing term originating from the reflected electrons~\cite{ali2017}.
To find the distribution function in the detector, we insert $h_j = h_0 + \v h\cdot \v \sigma$ and $h_i = h^\text{det}_0 + \v h^\text{det}\cdot \v \sigma$, where the former is the distribution function in the nanowire and the latter is the distribution function in the detector.
From this, we get
\begin{subequations}
\begin{align}
  N_0^\text{det}D^\text{det}\uv n \cdot \tilde\nabla\circ h^\text{det}_0 &= \zeta \left(h_0 - h^\text{det}_0\right) + P\zeta\left(h_z - h^\text{det}_z\right),
  \\
  N_0^\text{det}D^\text{det}\uv n \cdot \tilde\nabla\circ h^\text{det}_z &= \zeta \left(h_z - h^\text{det}_z\right) + P\zeta\left(h_0 - h^\text{det}_0\right),
  \\
  N_0^\text{det}D^\text{det}\uv n \cdot \tilde\nabla\circ h^\text{det}_x &= \zeta \left(\sqrt{1-P^2}h_x - h^\text{det}_x\right) - iP\zeta h_y^\text{det} + G_\phi h^\text{det}_y,
  \\
  N_0^\text{det}D^\text{det}\uv n \cdot \tilde\nabla\circ h^\text{det}_y &= \zeta \left(\sqrt{1-P^2}h_y - h^\text{det}_y\right) + iP\zeta h_x^\text{det} - G_\phi h^\text{det}_x.
\end{align}
\end{subequations}
We can see that the boundary conditions couple $h_0^\text{det}$ with $h_z^\text{det}$ and $h_x^\text{det}$ with $h_y^\text{det}$.
We can rewrite the boundary condition in terms of current densities, voltages and spin-voltages by multiplying with $\pi/\abs{e}$. where $e$ is the electron charge, Fourier transforming and letting $t\to 0$, giving
\begin{subequations}
\begin{align}
  -\uv n \cdot \v J_e  &= \frac{\zeta}{N_0^\text{det}}\left(P\left[\mu_z - \mu_z^\text{det}\right] - V^\text{det}\right),
  \label{eq:charge_curr_det}
  \\
  -\uv n \cdot \v J_z &= \frac{\zeta}{N_0^\text{det}}\left(\mu_z - \mu_z^\text{det} - PV^\text{det}\right),
  \label{eq:spin_curr_det}
\end{align}
\end{subequations}
where $J_e = -\lim_{t\to 0}(\pi D^\text{det}/\abs{e})\tilde\nabla\circ H_0^\text{det}$ and $J_z = -\lim_{t\to 0}(\pi D^\text{det}/\abs{e})\tilde\nabla\circ H_z^\text{det}$ is the normalized charge current density and spin current density in the $z$-direction, respectively.
The electrochemical potential is
\begin{equation}
  V^\text{det} = \frac{1}{2\abs{e}} \int_{-\infty}^\infty \dd{\varepsilon} \left[h_0^\text{det} - \tanh(\beta\varepsilon/2)\right].
\end{equation}
We have set the electrochemical potential on the nanowire side to be $0$.

To solve for $\mu_z^\text{det}$ and  $V^\text{det}$ we must use solve Usadel equation.
For concreteness we assume that the electrode is a normal metal, but the relevant equations will be the same if it was instead a ferromagnet that is weakly polarized in the $z$-direction.
For the case of a normal metal, the Usadel equation is
\begin{equation}
  \frac{\partial h^\text{det}}{\partial T} - D^\text{det} \tilde\nabla\circ\tilde\nabla\circ h^\text{det} + 2\delta (h^\text{det} - h_\text{eq}) + \frac i 2 \comm{\check \sigma_\text{sd}}{\check g_s^\text{det}}_\circ^K = 0,
  \label{eq:usadel_det}
\end{equation}
where $\check\sigma_\text{sp}$ is a source of spin-diffusion.
This could for instance come from scattering with magnetic impurities, in which case $\check \sigma_\text{sd} = -i\v n \cdot \v\sigma\check g_s^\text{det}\v n \cdot \v\sigma/2\tau_\text{sd}$, where $\tau_\text{sd}$ is the scattering time and $\v n$ is the magnetization direction of the magnetic impurities~\cite{hugdal2017}.
Spin-diffusion could also come from spin-orbit coupling, as is the case in the nanowire.
Here we assume that spin-diffusion come from magnetic impurities rather than SOC.
Since the spin-accumulation in the $z$-direction is static in the nanowire and since there is no coupling between $\mu^\text{det}_z$ and $\mu^\text{det}_x$ or $\mu^\text{det}_y$, we can look for static solutions to $\mu^\text{det}_z$ and $V^\text{det}$.
From \cref{eq:usadel_det} we find that these solve
\begin{subequations}
\begin{align}
  -D^\text{det}\nabla^2 V^\text{det} = 0 = \nabla\cdot\v J_e,
  \label{eq:usadel_det_charge}
  \\
  -D^\text{det}\nabla^2 \mu_z^\text{det} + 2\left(\delta + \frac{1}{\tau_{\text{sd},x}} + \frac{1}{\tau_{\text{sd},y}}\right)\mu_z^\text{det}= 0,
  \label{eq:usadel_det_spin}
\end{align}
\end{subequations}
where $\tau_{\text{sd},x}$ and $\tau_{\text{sd},y}$ is the spin-diffusion times for magnetic impurities with magnetization in the $x$- and $y$-direction, respectively.
\Cref{eq:usadel_det_charge} states that charge is conserved inside the detector, so the electrical current is constant.
Since the detector is assumed to be an open circuit, $\uv n \cdot  \v J_e = 0$ on the far side of the detector, so from \cref{eq:charge_curr_det} we get that
\begin{equation}
  V^\text{det} = P\left[\mu_z - \mu_z^\text{det}\right].
  \label{eq:volt_1}
\end{equation}
From \cref{eq:usadel_det_spin} we see that the spin-accumulation decay exponentially inside the detector over a length scale given by the spin-diffusion time.
If we assume that the detector is thin we can approximate it as one-dimensional.
Let the length of the detector be $L^\text{det}$ and the axial coordinate be $s$, then
\begin{equation}
  \mu_z^\text{det} = C\cosh(k_z \left[L^\text{det} - s\right]),
\end{equation}
where $s = 0$ is at the interface with the nanowire, $C$ is a constant and
\begin{equation}
  k_z = \sqrt{2\left(\delta + \frac{1}{\tau_{\text{sd},x}} + \frac{1}{\tau_{\text{sd},y}}\right)/D^\text{det}}
\end{equation}
is the inverse spin-diffusion length.
The coefficient $C$ can be found from \cref{eq:spin_curr_det}.
From this we find that the spin-voltage in the detector at the interface is
\begin{equation}
  \mu^\text{det}_z = \frac{\zeta(1-P^2) \mu_z}{N_0^\text{det}D^\text{det}k_z\tanh(k_z L^\text{det}) + \zeta(1-P^2)}.
\end{equation}
We see that $\abs{\mu_z^\text{det}} \ll \abs{\mu_z}$ if the spin-diffusion length, $1/k_z$, or interface conductance, $\zeta$, is sufficiently small or the polarization, $P$, is sufficiently close to 1.
In particular, $\abs{\mu_z^\text{det}} \ll \abs{\mu_z}$ if 
\begin{equation}
  \frac{1}{k_z} \ll \frac{N_0^\text{det}D^\text{det}\tanh(k_z L^\text{det})}{\zeta(1-P^2)}.
\end{equation}
In this case \cref{eq:volt_1} reduces to
\begin{equation}
  V^\text{det} = P\mu_z,
\end{equation}
which is the same as \cref{eq:curr_det_1}.
Since there is no current inside the detector, the electrochemical potential is constant and the voltage difference measured between the detector and nanowire will be $V = V^\text{det}$.

Finally, note that there is a finite spin-current into the detector if the polarization is different from $1$.
That is, the detector acts as a spin-sink.
This can affect the magnetization in the nanowire where the spin-voltage is supposed to be measured.
However, this effect can be neglected if the interface conductivity and contact area are small.

% Fakesection: Bibliogaphy
% \clearpage
\bibliography{bibliography}